# Meson-meson scattering in lattice QED$_{2+1}$ *

H. R. Fiebig[a] and R. M. Woloshyn[b]

[a]Physics Department, FIU University Park, Miami, Florida 33199, U.S.A.

[b]TRIUMF, 4004 Wesbrook Mall, Vancouver, B.C., Canada V6T 2A3

Scattering phase shifts of a meson-meson system in staggered 3-dimensional lattice QED are computed. The main task of the simulation is to obtain a discrete set of two-body energy levels. These are extracted from a 4-point time correlation matrix and then used to obtain scattering phase shifts. The results for the $\ell = 0$ and $\ell = 2$ partial waves are consistent with short-range repulsion and intermediate-range attraction of the residual meson-meson interaction.

## 1. INTRODUCTION

The residual interaction between two hadron-like particles is of fundamental interest in strong-interaction physics. In a finite-volume lattice simulation the interaction manifests itself in a discrete set of energy levels of the relative motion of the two particles. According to a recent proposal by Lüscher [1] those energy levels can be used to extract scattering phase shifts.

This program was implemented for a meson-meson system in lattice QED$_3$. While this model is computationally much less demanding than lattice QCD$_4$, it does have a confining phase and also exhibits chiral symmetry breaking. The shared phenomenology thus makes lattice QED$_3$ a suitable model for the problem at hand.

## 2. LATTICE SIMULATION

The sole purpose of the lattice simulation is to obtain the above-mentioned set of energy levels for the relative degrees of freedom.

### 2.1. Lattice model

We have used the compact version of the Wilson action

$$S_W = \beta \sum_\Box (1 - \cos \Theta_\Box) \qquad (1)$$

*This work was supported in part by the Supercomputing Research Institute SCRI at Florida State University and by the Natural Sciences and Engineering Research Council of Canada.

where $\Theta_\Box$ denotes the plaquette angle. In three dimensions the staggered fermion action

$$S_F = \sum_{x,y} \overline{\chi}_f(x) G^{-1}_{x,y} \chi_f(y) \qquad (2)$$

describes two fermion species [2]. An external flavour index $f = u, d$ was added to the Grassmann fields. The usual staggered fermion matrix is called $G^{-1}$ here.

In order to match the phenomenology of QCD$_4$ it is essential that the field is in the confined phase. On a $24^2 \times 32$ lattice, which was used throughout the simulation, this is guaranteed at $\beta = 1.5$ with $<W> = 0.03$ being the value for the Wilson line.

A second requirement is that the lattice be large enough to accommodate two mesons. To estimate the size of one meson, its mass $m$ was obtained from simulations on $L^2 \times 32$ lattices with various sizes $L$. The mass $m(L)$ saturates exponentially at large $L$ and is already within 5% of its asymptotic value at $L = 10$.

A total of 64 gauge field configurations was generated in the quenched approximation.

### 2.2. Meson field operators

A simple local one-meson operator is

$$\phi_{\vec{p}}(t) = L^{-2} \sum_{\vec{x}} e^{i\vec{p}\cdot\vec{x}} \overline{\chi}_d(\vec{x},t) \chi_u(\vec{x},t) \qquad (3)$$

where $\vec{p}$ belongs to the set of $L^2$ discrete lattice momenta. Two mesons at center-of-mass momentum $\vec{P} = 0$ are then described by the operator

$$\Phi_{\vec{p}}(t) = \phi_{-\vec{p}}(t) \phi_{\vec{p}}(t) = \Phi^\dagger_{\vec{p}}(t) \qquad (4)$$

### 2.3. Time-correlation functions

The threshold energy $2m$ of the noninteracting two-meson system is related to the square of the 2-point correlation function

$$C^{(2)}(t,t_0) = \qquad\qquad (5)$$
$$<\phi_{\vec{0}}^\dagger(t)\phi_{\vec{0}}(t_0)> - <\phi_{\vec{0}}^\dagger(t)><\phi_{\vec{0}}(t_0)>$$

The interacting two-meson system is described by the 4-point correlation matrix

$$C_{\vec{p}\vec{q}}^{(4)}(t,t_0) = \qquad\qquad (6)$$
$$<\Phi_{\vec{p}}^\dagger(t)\Phi_{\vec{q}}(t_0)> - <\Phi_{\vec{p}}^\dagger(t)><\Phi_{\vec{q}}(t_0)>$$

Both correlators $C^{(2)}$ and $C^{(4)}$ can be worked out in terms of contractions between the Grassmann fields contained in $\phi$ and $\Phi$. The simplistic flavour structure in (3) greatly reduces the number of terms in (6). There remain three basic diagrams that contribute to the 4-point correlation matrix

$$C^{(4)} = C^{(4A)} + C^{4B} - 2\operatorname{Re} C^{(4C)} \qquad (7)$$

For example

$$C_{\vec{p}\vec{q}}^{(4C)}(t,t_0) = \qquad\qquad (8)$$
$$L^{-8}\sum_{\vec{x}_1}\sum_{\vec{x}_2}\sum_{\vec{y}_1}\sum_{\vec{y}_2} e^{i\vec{p}\cdot(\vec{x}_2-\vec{x}_1)+i\vec{q}\cdot(\vec{y}_2-\vec{y}_1)}$$
$$G^*_{\vec{x}_2t,\vec{y}_1t_0}G_{\vec{x}_2t,\vec{y}_2t_0}G^*_{\vec{x}_1t,\vec{y}_2t_0}G_{\vec{x}_1t,\vec{y}_1t_0}$$

This expression (and the two similar ones for $C^{(4A)}$ and $C^{(4B)}$) poses serious computational problems for two reasons. First, the entire propagator matrix $G$ is needed. Second, the site sums have $L^8 \cong 10^{11}$ terms on the $L=24$ lattice. They are numerically intractable.

### 2.4. Random-source technique

The matrix elements of $C^{(4)}$ can be computed efficiently with Fourier-modified random sources. Let $R_{\vec{x}}$ be complex Gaussian random vectors, then define

$$H_{\vec{x}t,t_0}(\vec{p};R) = L^{-2}\sum_{\vec{y}} G_{\vec{x}t,\vec{y}t_0} e^{i\vec{p}\cdot\vec{y}} R_{\vec{y}} \qquad (9)$$

For each $\vec{p}$ and $R$ the vector $H$ can be obtained from a single call to, say, a conjugate-gradient routine. The 2-point and the 4-point correlators both have representations in terms of the auxiliary vectors (9). For example

$$C_{\vec{p}\vec{q}}^{(4C)}(t,t_0) = <\sum_{<R^{(1)}>}\sum_{<R^{(2)}>} \qquad (10)$$
$$\sum_{\vec{x}_1} e^{-i\vec{p}\cdot\vec{x}_1} H_{\vec{x}_1t,t_0'}(\vec{0};R^{(1)})^* H_{\vec{x}_1t,t_0}(\vec{0};R^{(2)})$$
$$\sum_{\vec{x}_2} e^{i\vec{p}\cdot\vec{x}_2} H_{\vec{x}_2t,t_0}(\vec{q};R^{(2)})^* H_{\vec{x}_2t,t_0}(\vec{q};R^{(1)})>$$

All contributions to $C^{(4)}$ require two independent random sources. The sums over $R^{(1)}$ and $R^{(2)}$ in (10) denote averages. For each average, 8 samples were used in the present simulation.

### 2.5. Properties of the correlation matrix

The eigenvalues $\lambda_n(t,t_0)$ of the 4-point correlation matrix determine the two-body energy levels $W_n$. We have [3]

$$\lambda_n(t,t_0) \cong c_n \cosh[W_n(t-t_c)] \qquad (11)$$

On the $24^2 \times 32$ lattice with periodic boundary conditions $t_c = 17$. Figure 1 shows the time dependence of the five eigenvalues with the largest coefficients $c_n$ together with cosh fits according to (11). The matrix $C^{(4)}$ was truncated to momentum indices $\vec{p} = (n_1,n_2)2\pi/L$ with $|n_{1,2}| \leq 2$. Certain symmetries of $C^{(4)}$ allow to restrict the range of momentum indices further to 13.

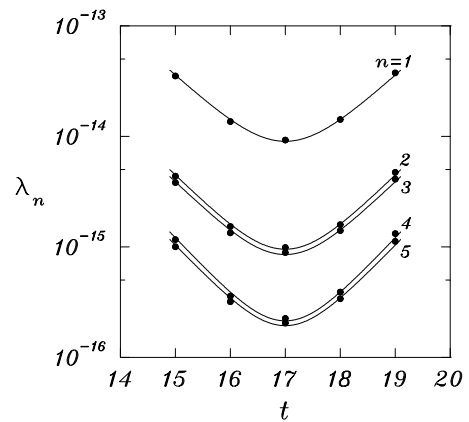

Figure 1. Eigenvalues of the 4-point correlator.

Table 1
Two-body strength factors and energy levels ($W_0 = 2m$).

| $n$ | $c_n[5 \times 10^{-15}]$ | $W_n$ | $k_n^2$ |
|---|---|---|---|
| 0 | 0.958(25) | 1.032(08) | 0.0 |
| 1 | 1.805(51) | 1.029(09) | |
| 2 | 0.191(06) | 1.114(09) | 0.047(3) |
| 3 | 0.171(05) | 1.098(10) | 0.038(3) |
| 4 | 0.043(01) | 1.209(12) | 0.117(6) |
| 5 | 0.039(01) | 1.181(13) | 0.095(6) |

## 2.6. Two-body energy levels

Table 1 shows the results of two-parameter fits to the eigenvalues according to (11). The threshold energy $W_0 = 2m$ was obtained from a fit to the square $C^{(2)}(t,t_0)^2$ of the 2-point correlator.

In principle, the relative energies

$$k^2 = \frac{1}{4}[W^2 - (2m)^2] \qquad (12)$$

could now be compiled from the $W_n$. In order to have better control over the errors, however, it is preferable to extract the $k_n^2, n = 1...5$, directly from fits to the ratios

$$R_n(t,t_0) = \lambda_n(t,t_0)/C^{(2)}(t,t_0)^2 \qquad (13)$$

The time dependence of the $R_n(t,t_0)$ is easily worked out using (11). With $W_n$ and $c_n$ fixed from the previous 2-parameter fits, the energies $k_n^2$ were obtained from 1-parameter fits to the ratios $R_n(t,t_0)$. The results are also shown in Table 1. The present data did not allow to make a fit to the $n = 1$ level.

Comparison of $W_0$ with $W_1$ in Table 1 suggest the possibility that the $n = 1$ level may be a two-meson bound state. Although the present data are inconclusive in this regard, the residual interaction clearly seems to contain an attractive component.

## 3. FINITE-VOLUME SCATTERING

The discrete set of energy levels $k_n^2$ of the relative motion, as listed in Table 1, is the basis for computing scattering phase shifts. Following Lüscher [1] a nonlinear relation between the continuum phase shifts in partial wave $\ell = 0, 1...\Lambda$ can be derived by considering periodic replications of the interaction potential and then match the physical (regular) solutions to the free (singular periodic) solutions of the Helmholtz equation. Although the general outline of [1] remains unchanged, nearly all details have to be recalculated for the $d = 2$ dimensional case [4].

### 3.1. Cubic symmetry

In $d = 2$ space dimensions the cubic symmetry group $O(2,Z)$ has 5 irreducible representations $\Gamma = A_1, A_2, B_1, B_2, E$ which are one dimensional with the exception of $E$, which has dimension two. The spaces $\underline{\ell}$ spanned by $Y_{\ell m}(\phi) \propto e^{im\ell\phi}$, with $m = 0$ for $\ell = 0$ and $m = \pm 1$ for $\ell = 1, ...\Lambda$, have the following decompositions

$$\left.\begin{array}{l}\underline{0} = A_1 \\ \underline{1} = E \\ \underline{2} = B_1 \oplus B_2 \\ \underline{3} = E^* \\ \underline{4} = A_1 \oplus A_2\end{array}\right\} \quad \text{modulo 4} . \qquad (14)$$

with $E$ and $E^*$ being equivalent.

### 3.2. Determinant condition

The above-mentioned nonlinear condition for the phase shifts reads

$$\prod_\Gamma \det\left[e^{2i\delta^{(\Gamma)}} - \frac{M^{(\Gamma)} - i}{M^{(\Gamma)} + i}\right] = 0 \qquad (15)$$

Here $M^{(\Gamma)} = M^{(\Gamma)}(q)$, with $q = kL/2\pi$, is a matrix of size $2\Lambda + 1$ (cut off at $\ell = \Lambda$) that contains the scattering information of the cubic $d = 2$ geometry. The (diagonal) matrix $e^{2i\delta^{(\Gamma)}}$ comprises the phase shifts $\delta_\ell^{(\Gamma)}(q)$ which are contained in the representation $\Gamma$, see (14).

The task now consists in solving (15) for each of the relative energies $q_n^2, n = 2...5$, listed in

Table 1. Clearly the solution is not unique. We have treated each of the five factors in (15) separately. Even then the solutions are not unique, in general. At an angular momentum cutoff of $\Lambda = 8$ the determinants have the sizes $3, 2, 2, 2, 4$ for $\Gamma = A_1, A_2, B_1, B_2, E$ respectively. The solutions, however, become unique if one assumes that within each space $\Gamma$ only one partial wave $\ell$ is dominant. This assumption [1] may be tested by setting all but one $\delta_\ell^{(\Gamma)}$ to zero. In this way one obtains 13 sets of phase shifts $\delta_\ell^{(\Gamma)}$.

## 4. THE PHASE SHIFTS

We find that most of these phase shifts are indeed small, as expected [1]. In particular all $\delta_\ell^{(\Gamma)}$ with $\ell$ odd are essentially zero. Significant structure is only found in the $A_1$-sector for $\ell = 0$ (s-wave) and the $B_2$-sector for $\ell = 2$ (d-wave). Those data are shown in Figures 2 and 3.

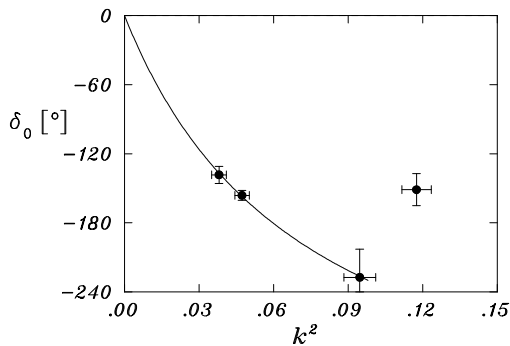

Figure 2. s-wave phase shifts ($\ell=0$) in sector $A_1$.

The monotonic decrease of the s-wave phase shift, up to about $k^2 \cong 0.1$, is consistent with a repulsive residual interaction. Since the meson-meson overlap at short distances is largest in the s-wave, this feature may simply indicate Pauli repulsion. The data point at $k^2 \cong 0.12$ appears to indicate some other type of (possibly lattice) effect, which is not yet understood.

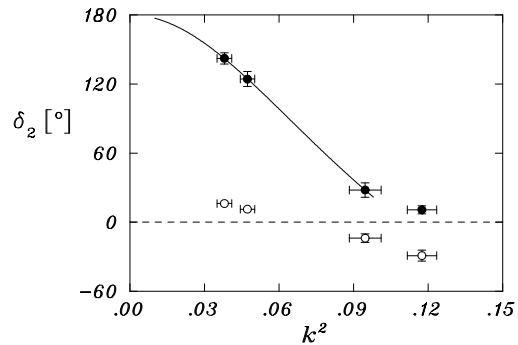

Figure 3. d-wave phase shifts ($\ell=2$) in sectors $B_2$ (●) and $B_1$ (○).

In the d-wave the data have been plotted such that the phase decreases by 180° as $k^2$ becomes large. This is consistent with either a bound state (Levinson's theorem) or a sharp resonance close to the threshold. In either case, since the d-wave overlap tests somewhat larger distances (possibly the edges of the mesons), this feature would indicate intermediate-range attraction of the residual interaction.

## 5. CONCLUSION

The results of the meson-meson scattering simulation in lattice $QED_3$ possibly suggest short-range repulsion and intermediate-range attraction of the residual force. These features allude to basic properties of hadronic interaction.